\documentclass[%
reprint,
superscriptaddress,
showpacs,
preprintnumbers,
amsmath,amssymb,
prl,
floatfix, ]%
{revtex4-1}

\usepackage{color}
\usepackage{CJK}

\usepackage{graphicx}
\usepackage{dcolumn}
\usepackage{bm}
\usepackage[dvipdfmx,bookmarks=true,colorlinks,%
            citecolor=blue,linkcolor=blue,anchorcolor=blue,filecolor=blue,urlcolor=blue,%
           ]{hyperref}

\begin{document}


\title{Non-Gaussian Fluctuation and Non-Markovian Effect in the Nuclear Fusion Process:
       Langevin Dynamics Emerging from Quantum Molecular Dynamics Simulations}
\author{Kai Wen}
 \affiliation{State Key Laboratory of Theoretical Physics,
              Institute of Theoretical Physics, Chinese Academy of Sciences,
              Beijing 100190, China}
\author{Fumihiko Sakata}
 \affiliation{Institute of Applied Beam Science, Graduate School of Science and Technology,
              Ibaraki University, Mito 310-8512, Japan}
 \affiliation{State Key Laboratory of Theoretical Physics,
              Institute of Theoretical Physics, Chinese Academy of Sciences,
              Beijing 100190, China}
\author{Zhu-Xia Li}
 \affiliation{China Institute of Atomic Energy, Beijing 102413, China}
\author{Xi-Zhen Wu}
 \affiliation{China Institute of Atomic Energy, Beijing 102413, China}
 \author{Ying-Xun Zhang}
 \affiliation{China Institute of Atomic Energy, Beijing 102413, China}
\author{Shan-Gui Zhou}
 \email{sgzhou@itp.ac.cn}
 \affiliation{State Key Laboratory of Theoretical Physics,
              Institute of Theoretical Physics, Chinese Academy of Sciences,
              Beijing 100190, China}
 \affiliation{Center of Theoretical Nuclear Physics, National Laboratory
              of Heavy Ion Accelerator, Lanzhou 730000, China}

\date{\today}


\begin{abstract}
Macroscopic parameters as well as precise information on the random force
characterizing the Langevin type description of the nuclear fusion process
around the Coulomb barrier
are extracted from the microscopic dynamics of individual nucleons by
exploiting the numerical simulation of the improved quantum molecular dynamics.
It turns out that the dissipation dynamics of the relative motion between
two fusing nuclei is caused by a non-Gaussian distribution of the random force.
We find that the friction coefficient as well as the time correlation function of
the random force takes particularly large values in a region a little bit
inside of the Coulomb barrier.
A clear non-Markovian effect is observed in the time correlation function
of the random force.
It is further shown that an emergent dynamics of the fusion process can be described
by the generalized Langevin equation with memory effects by appropriately
incorporating the microscopic information of individual nucleons through
the random force and its time correlation function.
\end{abstract}


\maketitle


The fusion of two nuclei is one of the major non-equilibrium processes in low
energy nuclear reactions where the fluctuation and dissipation play important roles.
It is rather difficult to describe the fusion process without significant simplifications.
Under various assumptions,
several macroscopic transport models have been introduced to evaluate the
formation of a compound nucleus in heavy-ion fusion reactions~\cite{%
Shen2002_PRC66-061602R,
*Zagrebaev2012_PRC85-014608,
*Aritomo2012_PRC85-044614,
*Siwek-Wilczynska2012_PRC86-014611,
*Liu2013_PRC87-034616,
Adamian1998_NPA633-409,
*Li2010_NPA834-353c,
*Gan2011_SciChinaPMA54S1-61,
*Nasirov2011_PRC84-044612,
*Wang2012_PRC85-041601R}.
However, the microscopic mechanism on how two colliding nuclei fuse,
especially how the relevant kinetic energy dissipates into the intrinsic degrees of freedom (DoF),
remains a subject requiring further research.

\label{modification:ab-initio_micro}
On the other hand, it is becoming feasible to get various 
information out of microscopic numerical simulations, like
time-dependent Hartree-Fock (TDHF) theories~\cite{Bonche1976_PRC13-1226,
Guo2007_PRC76-014601,*Guo2008_PRC77-041301R,
Washiyama2008_PRC78-024610, Washiyama2009_PRC79-024609,
Simenel2012_EPJA48-152},
the many-body correlation transport (MBCT) theory~\cite{Wang1985_AoP159-328},
the quantum molecular dynamics (QMD)~\cite{Aichelin1991_PR202-233},
the antisymmetrized molecular dynamics~\cite{Ono1999_PRC59-853},
and the fermion molecular dynamics~\cite{Feldmeier2000_RMP72-655}.
The TDHF theory is mainly based on the mean-field concept; in TDHF,
fluctuations of collective variables are considerably underestimated.
Much effort has been made to give a beyond-mean-field description of
fluctuations~\cite{Ayik2008_PLB658-174}. 
The $n$-body correlations are incorporated in the MBCT theory~\cite{Wang1985_AoP159-328}
which
has only been used in very light systems~\cite{Liu1996_NPA604-341}.

The QMD is a microscopic dynamical $n$-body theory which was successfully
used in intermediate-energy heavy-ion collisions (HIC)~\cite{Aichelin1991_PR202-233}.
An improved QMD (ImQMD) has been developed in order to extend the application
of QMD to low-energy HICs near the Coulomb
barrier~\cite{Wang2002_PRC65-064608,*Wang2004_PRC69-034608}.
A series of improvements were made in the ImQMD;
\label{modification:Pauli}
in particular, by using the phase space occupation constraint method~\cite{Papa2001_PRC64-024612}, 
the fermionic properties of nucleons is remedied, which is important for low-energy collisions.
Making full use of the microscopic information provided by ImQMD simulations,
in this Letter, we try to understand how the macroscopic fusion dynamics emerges
out of the microscopic one.

We focus on a simplest case of symmetric fusion process with the impact parameter equal to zero.
In this case, \label{modification:division}
the system can be divided into the left- and right-half parts
instead of a projectile and a target~\cite{Ayik2009_PRC79-054606}.
The relative motion between two centers of mass (CoM) of the left and right parts
is chosen as the relevant DoF to be described by the Langevin equation.
Our analysis is limited in a stage where the relative
distance ${R}$ is much larger than its width.

The one-dimensional generalized Langevin equation with memory effects
reads~\cite{Mori1965_PTP33-423, Sakata2011_PTP125-359, Frobrich1998_PR292-131}
\begin{eqnarray}
\frac{du(t)}{dt}
    =-\int_{-\infty}^t \gamma (t-t')u(t')dt'
     + \frac{1}{\mu}{\delta F}(t)
     -\frac{1}{\mu}\frac{dV({R})}{{d{R}}},
~\label{Lange}
\end{eqnarray}
where $u(t)$ is the relative velocity between the two parts,
{$\delta F(t)$} the random force felt by either part,
$\mu$ the reduced mass of the system,
$\gamma(t-t')$ the friction kernel,
and $V(R)$ the potential for the relative motion.

In the ImQMD model~\cite{Wang2002_PRC65-064608, Wang2004_PRC69-034608},
a trial wave function is restricted within
a parameter space $\{\textbf{r}_{j}, \textbf{p}_{j}\}$, where
$\textbf{r}_{j}$ and $\textbf{p}_{j}$ are mean values
of position and momentum operators of the $j$th nucleon
which is expressed by a Gaussian wave packet.
The time evolution of the trial wave function under an effective potential
is governed by the time-dependent
variational principle~\cite{Aichelin1991_PR202-233,Ono1999_PRC59-853,Feldmeier2000_RMP72-655}.
An expectation value of the Hamiltonian is given by using
an improved Skyrme potential energy density functional.
In this Letter, we concentrate on head-on collisions of $^{90}$Zr+$^{90}$Zr.
Ten thousand collision events were simulated.
Each simulation is started at ${R}={R}_0=30$ fm
and with an incident energy $E=195$ MeV. 
Numerical details
can be found in Refs.~\cite{Tian2008_PRC77-064603,*Zhao2009_PRC79-024614}.

\begin{figure}
\begin{centering}
\includegraphics[width=0.75\columnwidth]{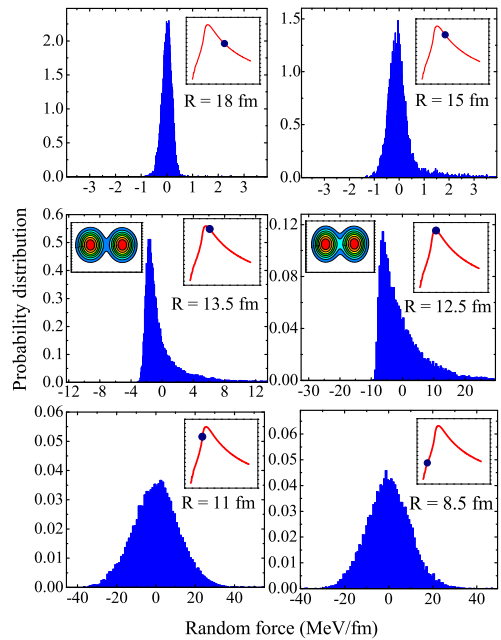}
\par\end{centering}
\caption{\label{fig:grows}(color online)
Distributions of the random force ${\delta F}({R})$. 
Each inset shows the potential $V({R})$ 
with the blue dot representing the position where the system locates.
The contour plots display the nucleon density distribution of the system.
}
\end{figure}

The potential for the relative motion is defined as,
\begin{eqnarray}
 V({R}) = E_{\mathrm{tot}}({R}) - E_{\mathrm{left}}({R}) - E_{\mathrm{right}}({R}),
~\label{eq:drive}
\end{eqnarray}
where 
$E_{\mathrm{tot}}$,
$E_{\mathrm{left}}$, and $E_{\mathrm{right}}$ represent the
energy of the system and those of the left and right
parts, respectively; each of which consists of the kinetic
energy, the nuclear and the Coulomb potential energies.
The potential $V(R)$ is shown in Figs.~\ref{fig:grows} and \ref{fig:fluc}.
The TDHF has also been used to extract microscopic interaction potentials between
two nuclei~\cite{Umar2006_PRC74-061601R,Washiyama2008_PRC78-024610}
which show similar features as those from the ImQMD
simulations presented here and in Refs.~\cite{Jiang2010_PRC81-044602,*Zanganeh2012_PRC85-034601}.

The random force or the fluctuation of force in the $i$th event is defined as
\begin{eqnarray}
 {\delta F(x)}_{i} & \equiv & F_{i}(x) - \langle F(x) \rangle, \ \ 
 x = t\ \mathrm{or}\ {R} ,
 \label{eq3}
\end{eqnarray}
where $F_{i}(x) \equiv \sum_{j=1}^{A} f^{j}_i(x)$ denotes the total force acting
on the left (right) part of the system in the $i$th event,
$\langle F(x) \rangle \equiv \frac{1}{n} \sum_{i=1}^{n} F_{i}(x)$
the mean value,
and $f^{j}_i(x)$ the force on the $j$th nucleon in the left (right) part.
Here $A$ means the number of nucleons contained in the left (right) part
and $n$ denotes the total number of events.
In Eq.~(\ref{eq3}) and hereafter, $\langle {Q} \rangle$ denotes
an average of ${Q}$ over all events. 
\label{modification:origin_fluc}
For low-energy collisions, the fluctuation mainly stems from the
initialization of each event in which the position and the momentum of
each particle are chosen randomly under certain conditions. 
With time this initial fluctuation propagates and is not smoothed out because
in QMD a many-body rather than a mean field problem is solved~\cite{Aichelin1991_PR202-233}.
 
Distributions of ${\delta F}({R})$ at several distances 
are shown in Fig.~\ref{fig:grows}.
\label{modification:FWHM}
The random force at ${R}(t=0)={R}_0$ shows a Gaussian distribution with
the full width at half maximum (FWHM) $\Gamma \approx $ 0.1 MeV/fm
which could be understood analytically as only the
Coulomb field is felt by the particles.
In a region far away from the barrier, e.g., ${R}\approx 18$ fm,
${\delta F}$ has a Gaussian distribution with $\Gamma \approx 0.5$ MeV/fm.
From a certain distance, ${R}\approx 13.5$ fm, there appears a non-Gaussian
shape, as is observed in Fig.~\ref{fig:grows}.
According to the shape of the distribution of ${\delta F}({R})$,
one may divide the whole process into three regions.
Region 1 represents an approaching phase up to the touching point:
The distribution has a Gaussian form with a rather narrow width.
Region 2 is from the touching point to the barrier top:
A non-Gaussian shape appears.
Region 3 is from just inside the barrier top to the fusing phase:
The distribution of ${\delta F}({R})$ has again a Gaussian shape
with $\Gamma \approx 15$ MeV/fm which is
almost two orders of magnitude larger than that in Region 1.
 
\begin{figure}
\begin{centering}
\includegraphics[width=0.75\columnwidth]{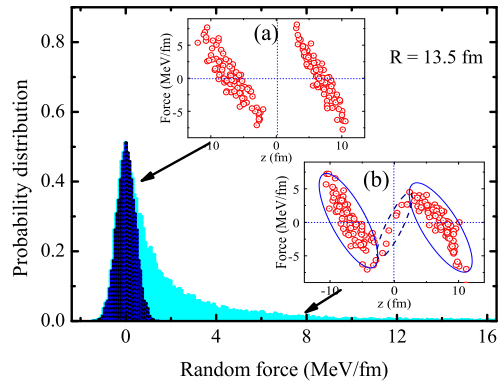}
\par\end{centering}
\caption{\label{fig:asymm}(color online)
Distribution of the random force ${\delta F}({R})$ at ${R} =$~13.5~fm
which is divided into the symmetric Gaussian (dark blue)
and asymmetric tail (light blue) parts.
Two typical events are shown in the inset:
The abscissa and the ordinate express relative position $z$ of each nucleon and
the force it feels in the $z$ direction.
}
\end{figure}
%
To make clear what happens in Region 2,
we divide the distribution of ${\delta F}({R})$ into a symmetric Gaussian and
an asymmetric tail parts as is shown in Fig~\ref{fig:asymm}.
The width of the Gaussian part is of the same order of magnitude as that in Region 1.
The detailed structure of the random force can be studied by examining
the strength and direction of the force felt by each nucleon.
One typical event in the symmetric part is shown in
Fig.~\ref{fig:asymm}(a): All nucleons are well divided
into two separated groups expressing the projectile and the target, respectively.
Moreover, each nucleon locating in the left
side of each nucleus feels a force toward the right (positive value),
and that in the right side feels a force toward the left
(negative value), so as to keep a stable mean-field.
The resultant force made by all nucleons in each nucleus is almost zero.
Namely, the intrinsic structure of two fusing
nuclei is kept almost unchanged, so is the width of the random force.
This situation persists in events which belong to the symmetric Gaussian part
in Region 2 and in all those in Region 1.

A typical event in the asymmetric tail is shown in Fig.~\ref{fig:asymm}(b).
Nucleons are roughly divided into two groups surrounded by solid lines.
However, there appears a small third group within the dashed line.
Since a few points in the negative (positive) force region express a set of
nucleons which escape from the left (right) nucleus,
and are being absorbed by the right (left) nucleus,
a resultant force made by these nucleons gives
a large right(left)-directed component to the random force.
These transferred nucleons
move in an average potential formed by both the projectile and the target;
they play a role to open a {\it window}.


When the two nuclei come much closer, there occur more events
which have more nucleons in the third group.
Meanwhile, the other two groups, originating from the projectile and target,
become closer to each other.
Consequently, the asymmetric tail in the distribution of
${\delta F}({R})$ becomes larger.
At the border between Regions 2 and 3, it becomes very difficult to distinguish
an event in the center part of the distribution from that in the tail part
and all events are absorbed into a widely spreading Gaussian distribution.

From above discussions, it is concluded that the main microscopic origin
of the random force, i.e., a two orders of magnitude enhancement of
the random force is generated by individual nucleons
in the third group.
These nucleons also result in the abnormal behavior in
the distribution of ${\delta F}({R})$, i.e., the long tail in Region 2 and
a much larger width in Region 3 compared to Region 1.

Next let us extract information for the macroscopic
dynamics out of microscopic simulations.
Assuming that the work done by the friction force is completely
converted into the intrinsic energy $E_{\mathrm{intr}}({R}) $,
\label{modification:Rayleigh}
one gets the friction coefficient $\gamma_0(R)$ from the Rayleigh 
formula~\cite{Washiyama2009_PRC79-024609, Ayik2009_PRC79-054606, Frobrich1998_PR292-131},
\begin{equation}
 \gamma_0({R}) \equiv \frac{ \langle F_{\mathrm{fric}}({R})\rangle}{\langle P\rangle _{{R}}} ,
~\label{Eq:fric}
\end{equation}
with
$F_{\mathrm{fric}}({R}) \equiv  dE_{\mathrm{intr}}({R})/d{R}$,
$E_{\mathrm{intr}}({R}) \equiv  E_{\mathrm{tot}}({R}) - E_{\mathrm{coll}}({R})$,
and $E_{\mathrm{coll}}({R})={P^2}/{2\mu }+V({R})$.
$P$ denotes the relative momentum between two CoMs
and its mean value $\langle P\rangle_{{R}}$ at a given ${R}$ is defined as
\begin{equation}
 \langle P \rangle_{{R}} \equiv
 \frac{1}{n} \sum_{i=1}^n \left. P_i(t_i) \right|_{\{ t_i|{R}_i(t_i)={R}\}},
\label{eq:p_mean}
\end{equation}
where $P_i(t)$ and ${R}_i(t)$ are the momentum and coordinate of the $i$-th event at time $t$
and the following correspondence is used: 
For each event $i$, a time $t_i$ is chosen in such a way that the
relative distance takes a given value ${R}$, i.e., ${R}_i(t_i)={R}$.
A ${R}$-dependent correlation function is defined as
\begin{equation}
 \langle {\delta F}({R}){\delta F}({R})\rangle
 \equiv
 \frac{1}{n} \sum_{i=1}^n \left. {\delta F}_i(t_i){\delta F}_i(t_i) 
                          \right|_{\{t_i|{R}_i(t_i)={R}\}}
 .
 \label{eq:correlation_r}
\end{equation}
\label{modification:t_R}
 
\begin{figure}
\begin{centering}
\includegraphics[width=0.88\columnwidth]{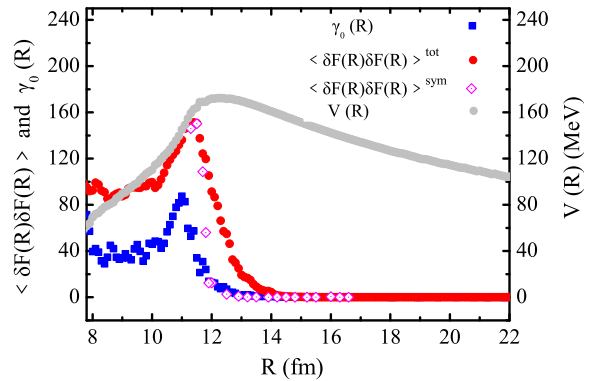}
\par\end{centering}
\caption{~\label{fig:fluc}(color online)
The correlation function $\langle {\delta F}({R})
{\delta F}({R})\rangle^\mathrm{tot}$ [red dots, in (MeV/fm)$^{2}$]
and the friction coefficient $\gamma_0({R})$ (blue squares, in
$0.001~c$/fm).
The grey line shows the potential $V({R})$.
Pink diamonds represent $\langle {\delta F}
({R}){\delta F}({R})\rangle^\mathrm{sym}$ calculated by eliminating
events in the asymmetric tail.
}
\end{figure}
Figure~\ref{fig:fluc} shows the correlation function $\langle 
{\delta F}({R}){\delta F}({R})\rangle$
and the friction coefficient $\gamma_0({R})$ which play decisive roles
in the macroscopic description of dissipation phenomena.
As is seen from Fig.~\ref{fig:fluc}, $\langle {\delta F}({R}){\delta F}({R})\rangle$
and $\gamma_0({R})$ have similar shapes and their peaks locate at similar ${R}$.
The friction coefficient of the fusion process induced by a head-on collision
extracted from TDHF calculations shows similar strong peak structure.
\label{modification:E_increase_1}
As the incident energy $E$ increases, the shape of the curve 
$\gamma_0(R)\sim R$ may change. When $E$ is high enough, 
$\gamma_0(R)$ increases gradually with decreasing 
$R$~\cite{Washiyama2009_PRC79-024609,Ayik2009_PRC79-054606,Wen2013_in-prep}. 
\label{modification:just}

\begin{figure}[t]
\begin{centering}
\includegraphics[width=0.8\columnwidth]{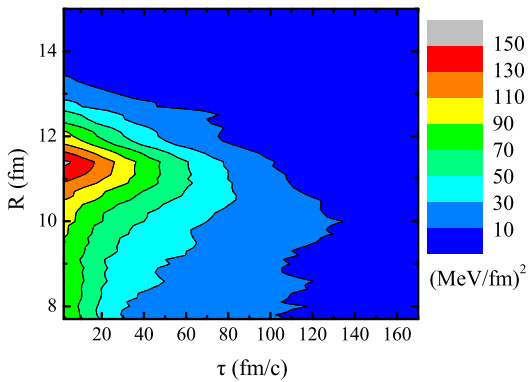}
\par\end{centering}
\caption{\label{fig:corr}(color online)
Time correlation function $\sigma(r,\tau)$~(\ref{eq:correlation_t}).
}
\end{figure}

To explore more deeply the dynamical relation between
the microscopic motion of individual nucleons and the macroscopic dissipative motion,
in Fig.~\ref{fig:corr} we show
the time correlation function of the random force ${\sigma}({R},\tau)$
which is defined as
\begin{eqnarray}
 {\sigma}({R},\tau) & \equiv &
 \frac{1}{n} \sum_{i=1}^n \left. {\delta F}_i(t_i) {\delta F}_i(t_i-\tau ) \right|_{\{t_i|{R}(t_i)={R} \}}.
 \label{eq:correlation_t}
\end{eqnarray}
\label{modification:App}
 
In Fig.~\ref{fig:corr} one clearly finds the non-Markovian effect.
Especially when ${R}=12\sim10$ fm, it is important to take account of
memory effects generated by the microscopic motion of nucleons
when one tries to properly evaluates macroscopic effects of the dissipation.
Starting from the generalized Langevin equation (\ref{Lange})
with memory effects, one gets a generalized fluctuation-dissipation (GFD) relation
$\langle {\delta F}(t){\delta F}(t-t')\rangle = \mu T \gamma (t-t')$
which properly takes account of the time correlation of the random force.
There are many ways to define the temperature for compound nuclei
(see, e.g., Ref.~\cite{Su2012_PRC85-017604}).
Here we define an effective temperature for colliding systems
by applying the GFD relation, 
\begin{eqnarray}
 T_{\mathrm{non-Markov}}(R)
 & = &
 \frac{1}{\mu \gamma_0(R)}
 \int_0^\infty d\tau {\sigma}({R},\tau) 
 .
 \label{eq:T}
\end{eqnarray}
 
\begin{figure}[t]
\begin{centering}
\includegraphics[width=0.80\columnwidth]{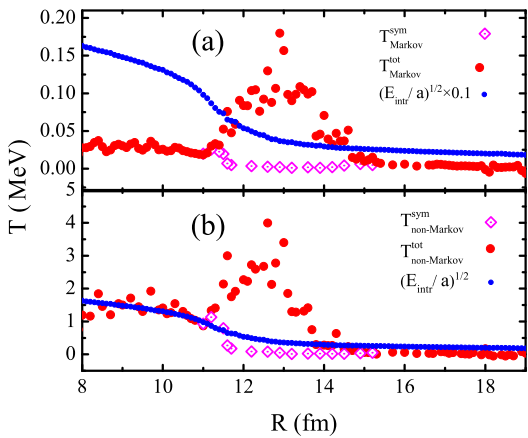}
\par\end{centering}
\caption{\label{fig:temperature2}(color online)
Effective temperature in the Markovian limit (a) and the Non-Markovian one (b). 
The blue line shows $\sqrt{E_{\textrm{intr}}/a}$ with $a = A_\mathrm{total}/4$.
}
\end{figure}
\label{modification:Fig.5_discussion}
The effective temperature $T_{\mathrm{non-Markov}}^{\textrm{tot}}$
as well as the one from the Markovian approximation $T_{\mathrm{Markov}}^{\textrm{tot}}$ 
are shown in Fig.~\ref{fig:temperature2}. 
$\sqrt{E_\mathrm{intr}/a}$ representing the temperature of a compound nucleus 
in the Fermi gas model is also shown as a reference.
Although $T_{\mathrm{non-Markov}}^{\textrm{tot}}$ and $T_{\mathrm{Markov}}^{\textrm{tot}}$
differ by one order of magnitude, they both
show a peak around the range where the asymmetric tail appears
in the distribution of ${\delta F}({R})$.
These peaks are related to the fact that the relative motion for events in 
the asymmetric tail part of the $\delta F(R)$ distribution
is strongly affected by a few transferred nucleons between two fusing
nuclei, i.e., by those in the third group of Fig.~\ref{fig:asymm}(b).
The macroscopic dynamics of the relative motion
described by the one-dimensional Langevin equation (\ref{Lange})
is not appropriate in Region 2.
In other words, the appearance of the non-Gaussian distributed random force
indicates a necessity of introducing a new macroscopic DoF.
Whether or not this new DoF may be related to the formation of a neck
is an open question~\cite{Siwek-Wilczynska2012_PRC86-014611,
*Zagrebaev2012_PRC85-014608,
*Shen2002_PRC66-061602R,
*Aritomo2012_PRC85-044614,
*Liu2013_PRC87-034616,
Adamian1997_NPA619-241,*Adamian2000_NPA671-233,*Diaz-Torres2000_PLB481-228}.

After eliminating the events in the asymmetric tail in the distribution of ${\delta F}({R})$,
one gets effective temperatures $T_{\mathrm{non-Markov}}^{\textrm{sym}}$
and $T_{\mathrm{Markov}}^{\textrm{sym}}$
which are depicted in Fig.~\ref{fig:temperature2}. 
\label{modification:fric_fluc}
The correlation function $\langle\delta F(R)\delta F(R)\rangle$ after eliminating
the asymmetric tail is also shown in {Fig.~\ref{fig:fluc}}.
$T_\mathrm{non-Markov}^\textrm{sym}$ shows a consistent feature
with $\sqrt{E_\mathrm{intr}/a}$ in Region 3.
While $T_\mathrm{Markov}^\textrm{sym}$ is by an order of magnitude
smaller than $T_\mathrm{non-Markov}^\textrm{sym}$.
\label{modification:ener_lost}
That is, the amount of energy dissipated from the relative motion
into the intrinsic DoFs could be more properly described
by the generalized Langevin equation with memory effects.

\label{modification:spin-orbit}
When the incident energy $E$ is far above the Coulomb barrier, 
the non-Gaussian fluctuation and the non-Markovian effect become less 
pronounced~\cite{Wen2013_in-prep}. 
It will be interesting to study the dependence of 
the non-Gaussian fluctuation and the non-Markovian effect
on $E$ as well as the impact parameter and the reaction system.
The spin-orbit coupling is important to properly reproduce 
the dissipation in heavy-ion fusion reactions~\cite{Umar1986_PRL56-2793}; 
e.g., the so-called ``fusion-window'' problem 
was solved in the first quantitative TDHF calculations with 
the inclusion of the spin-orbit interaction~\cite{Umar1986_PRL56-2793}.
One may expect more dissipations if the spin-orbit coupling effects are 
included in the ImQMD simulations.

\label{modification:around_VB}
In summary, we have discussed the generalized Langevin dynamics
with memory effects by using both the macroscopic and microscopic information
extracted from ImQMD simulations for the fusion process around the Coulomb barrier.
It is found that the dissipation dynamics of the relative motion between
two fusing nuclei is associated with non-Gaussian distributions of the random force.
In addition to the macroscopic information like the friction coefficient
and the potential for the relative motion,
the microscopic information of the random force as well as of its time correlation function
and a proper treatment of the non-Markovian (memory) effect in the Langevin dynamics
are decisive
for the dynamics of emergence in the nuclear dissipative fusion motion.

We thank G. Adamian, P. Danielewicz, Q. F. Li, B. N. Lu, R. Shi, S. J. Wang,
Y. T. Wang, Z. H. Zhang, E. G. Zhao, K. Zhao, and Y. Z. Zhuo for helpful discussions.
F.S. appreciates the support by
Chinese Academy of Sciences (CAS) Visiting Professorship for Senior
International Scientists (Grant No. 2011T1J27).
This work has been partly supported by
MOST of China (973 Program with Grant No. 2013CB834400),
NSF of China (Grants No. 11005155, No. 11075215, No. 11121403, No. 11120101005,
No. 11275052, and No. 11275248),
and Knowledge Innovation Project of CAS (Grant No. KJCX2-EW-N01).
The results described in this paper are obtained on the
ScGrid of Supercomputing Center,
Computer Network Information Center of CAS.

%


%

\end{document}